# Bi-serial DNA Encryption Algorithm (BDEA)

D.Prabhu * ,M.Adimoolam**
* Lecturer, Dept of Information Technology, Mailam Engineering College, Mailam
(Affiliated to Anna University of Technology, Chennai).
datatycoon@gmail.com
* Lecturer, Dept of Information Technology, Christ College of Engineering and Technology ,
(Affiliated to Pondicherry University , Puducherry).
m.adimoolam@gmail.com

**Abstract:** *The vast parallelism, exceptional energy efficiency and extraordinary information inherent in DNA molecules are being explored for computing, data storage and cryptography. DNA cryptography is a emerging field of cryptography. In this paper a novel encryption algorithm is devised based on number conversion, DNA digital coding, PCR amplification, which can effectively prevent attack. Data treatment is used to transform the plain text into cipher text which provides excellent security.*

## I. INTRODUCTION

In a pioneer study, Adleman[1] demonstrated the first DNA computing, which used DNA to solve a directed Hamiltonian path problem .It marked the beginning of a new stage in the era of information .This approach has been extended by Lipton[2] to solve another NP- complete problem which is the satisfaction problem .In the following researches, scientists fond that the vast parallelism, exceptional energy efficiency and extra ordinary information density are inherent in DNA molecules. DNA computing provided, a parallel processing capability with molecular level, introducing a new data structure and calculating method. It can simultaneously attack different parts of the computing problem put forward challenges and opportunities to traditional information security technology .For example in 1995,boneh et al[6]. demonstrated an approach to break the Data Encryption Standard (DES) by using DNA computing methods. In 1999 Clelland[10] et al. achieved an approach to Steganography by hiding secret message encoded as DNA standards among a multitude of random DNA.DNA is a new born cryptographic field emerged with the research of DNA computing. In which DNA is used as information carrier, this modern biological technology is used as an implementation tool. The vast parallelism and extra ordinary information density inherent in DNA molecules are explored for cryptographic purpose such as encryption, authentication, and signature and so on.

The research of DNA cryptography is still at the initial stage and there are many problem solved .The new born DNA cryptography is for from mature both in theory and realization and this might be the reason why only few example of DNA cryptography where proposed .On the other hand current DNA technology is beta stage of laboratory exploration and focus on experiments ,why there is not a general theory about applying DNA molecule in cryptography .Some key technologies in DNA research such as polymerase chain reaction (PCR).

DNA synthesis and DNA digital coding have only been developed and well accepted in recent years. PCR is a fast DNA amplification technology based Watson–crick complementary model, but it would still be extremely difficult to amplify the message –encoded sequence without knowing the correct two primer pairs. DNA cryptography should be implemented by using modern biological technique as tools and biological hard problems as main security basis to fully exert the special advantages .As described above, the correct two primer pairs would be used as the key by applying the special function of primers to PCR amplification. On the other hand the security of traditional cryptography is based on difficult mathematic problem which is mature both in theory and realization. There are many effective cryptosystem of traditional cryptography such as DES, AES, RSA are implemented widely some even have been extended (DDES & TDES). There by DNA cryptography does not absolutely repulse traditional cryptography and it is possible to construct hybrid cryptography

By studying the analysis of encryption technique, several algorithms are proposed on the basis of traditional encryption technique like DES, RSA, and AES etc. Most of our encryption are based on the basis of number system conversion .An analysis of algorithm of an encryption scheme using DNA technology, the message text will be converted into DNA digital code format, then PCR amplification is implemented there by creating large number of primes generation result large value of DNA data of the form (A, T, G, C).

Though of that, we have implemented the idea of collection of various process of number of conversion, DNA digital coding, PCR amplification, Data compression etc.

## II. DIFFICULT BIOLOGICAL PROBLEM USED IN THIS SCHEME

DNA is the germ plasm of all life styles. In a double helix DNA string, two strands are complementary in terms of sequences, that is A to T and C to G according to Watson-crick rules, which is one of the greatest scientific discoveries.

The modern cryptography is based on difficult mathematic problem such as the Non-Deterministic Polynomial Time Complete(NP-C)Problem. Quantum cryptography is based on Heinsberg uncertain principle, which can also be regarded as a difficult biological problem.

Some unresolved difficult biological problem in DNA science might have special value in cryptography and can achieve a new encryption technique. There are more difficult problem are more complex than biological problem. This absolutely different from well studied difficult mathematical problems. Here in our study we selected a typical difficult biological problem to develop an encryption scheme and tried to discuss the security of this scheme.

The difficult biological problem referred as "It extremely difficult to amplify the message encoded sequence without knowing the correct PCR two primer pairs "Polymerase Chain Reaction(PCR) is fast DNA amplification technology base complementary and is one of the most invention in modern biology .Two complementary oligonucleotide primers are annealed to double-standard target DNA strands ,and necessary target DNA strands and the necessary target DNA can be amplified after a serial of polymerase enzyme.. PCR is a very sensitive method and a single DNA target molecule can be amplified into $10^{6}$ after 20 cycles in theory, this can done within short time. It is a special function in PCR amplification that having correct primer pairs. It would extremely difficult to amplify the message encoded sequence without knowing the correct the primer pairs. So believe that this biological problem is difficult and will last a relatively long time.

## III. XOR OPERATION

XOR operation is applied over the message and key to increase the repetition of normalized binary bit of 1's and 0's, there by gaining high compression factor.

| TRUTH TABLE | | |
|---|---|---|
| X | Y | OUTPUT |
| 0 | 0 | 0 |
| 0 | 1 | 1 |
| 1 | 0 | 1 |
| 1 | 1 | 0 |

*Table1: XOR operation*

## IV. DNA DIGITAL CODING TECHNOLOGY

In the information science , the most fundamental coding method is Binary Digital Coding, which is anything can be encoded by two state 0 or 1 and a combination of o and 1,there are four kind of bases, which are ADENINE(A) and THYMINE(T) or CYTOSINE(C) and GUANINE(G) in DNA sequence. The simplest coding pattern to encode nucleotide bases (A,T,G,C) is by means four digits: 0(00),1(01),2(10),3(11),there are possibly 4!=24 possible pattern by encoding format like (0123/CTAG).

| BINARY VALUE | DNA DIGITAL CODING |
|---|---|
| 00 | A |
| 01 | T |
| 10 | G |
| 11 | C |

Pros:

1. To decrease redundancy of information coding and increase the efficiency compared to traditional encoding methods.

2. By using the technology of DNA digital coding, the traditional encryption method such as DES or RSA could be used to preprocess the plaintext.

3. The digital coding of DNA sequence convenient for mathematical operation and logical operation.

## V. PROPOSED SYSTEM

In our system, get the text based message from the user and convert the message into Hexadecimal code and Binary code. Message is split into parts, one is used as message end other one is used as key, the XOR operation is performed for the purposed of high compression factor.

DNA digital coding is applied over the message and get the DNA base coded message, then PCR amplification is implemented by using two prime pair as key and compression is performed for variable length of data.

Though by various mode of operation happen in serial fashion and provide double layer (Bi-Layer)security and therefore it is called as Bi-serial DNA Encryption Algorithm(BDEA).

and PCR amplification is performed by using two primer pairs. Using the primer pairs, we can generated the maximum populated data depend upon the primer pairs and then high compression algorithm is used to compress the variable - length data to minimize the memory size.

The $k$A and $k$B size to be same, but key values are different from each other, after decryption process combining the correct keys only. we get the original data, otherwise there is a possibility of missing some part of data.

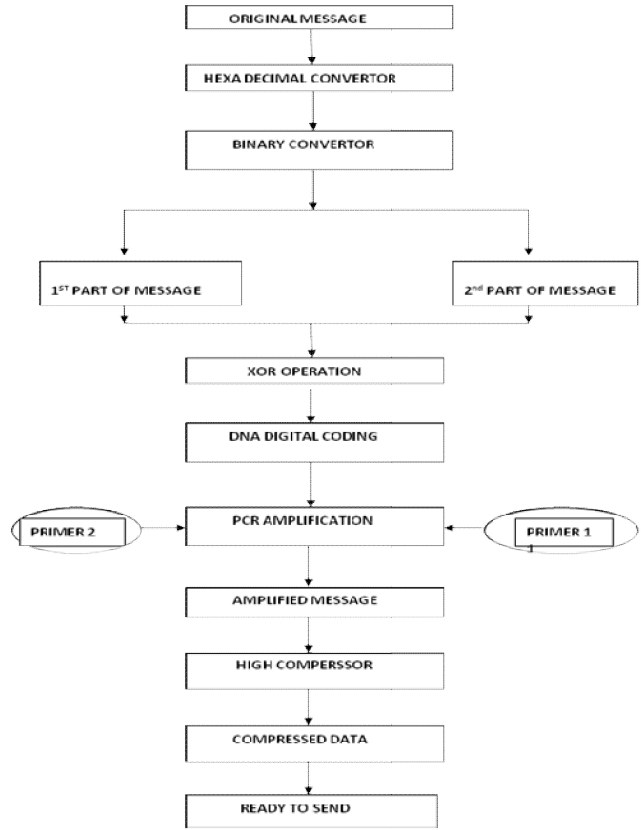

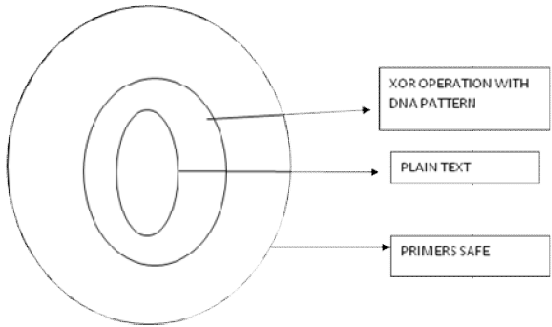

**Fig1:  Double Layer Security**

### VI. ENCRYPTION

A plain text is converted into Hexadecimal code and then converted into Binary code using decimal convertor. XOR operation can be performed using the code as Binary code , in that one part is used as a message($k$A) and another part used as key($k$B), using the Binary code XOR operation to be performed and the XORed output is taken for DNA digital coding to pre-process the binary text into DNA digital code

Fig2:ENCRYPTION

### VII. KEY GENERATION AND TRANSMISSION

Diffie–Hellman key exchange (D–H) is a cryptographic protocol that allows two parties that have no prior knowledge of each other to jointly establish a shared secret key over an insecure communications channel. This key can then be used to encrypt subsequent communications using a symmetric key cipher.

Generation of keys

1. Primer1.

2. Primer 2.

3. Hexadecimal key**.**

Three keys are send to the receiver using key agreement proposed by Diffie–Hellman.

## VIII. DECRYPTION

From the receiver side, get the encrypted data, by using high decompression algorithm to recover compressed data, then using the correct two primer pairs to retrieve the DNA digital coding.

DNA digital coding is converted into binary code , then XOR operation is performed by using the binary code and key is given by user, then combining the key and XORed output value to retrieve large Binary code and converted into Hexadecimal code. Hexadecimal code is converted into normal plain text by using Decimal Convertor.

In case anyone of the key is wrong, then there is possibi1lity missing of data or improper form of data. So, there is chance of maintaining more secure of data.

*EXAMPLE:*

**SAMPLE TEXT:**

CRYPTO

**HEXA VALUE:**

63727970746F

**BINARY VALUE:**

011000110111000100111100110111000001110100011 01111

**KEY VALUE:** 70746F

**DNA DIGITAL CODING:**

TGACTCAGTCGTTCAATCTATGCC

**PRIMER 1**: A

**PRIMER 2:** T

**FINAL AMPLIFIED CIPHER TEXT:**

TTATGTATCTATTTATCTATATATGTATTTATCTATGT
ATTTATTTATCTATATATATATTTATCTATTTATATAT
TTATGTATCTATCTAT

Similarly same method is used for Decryption.

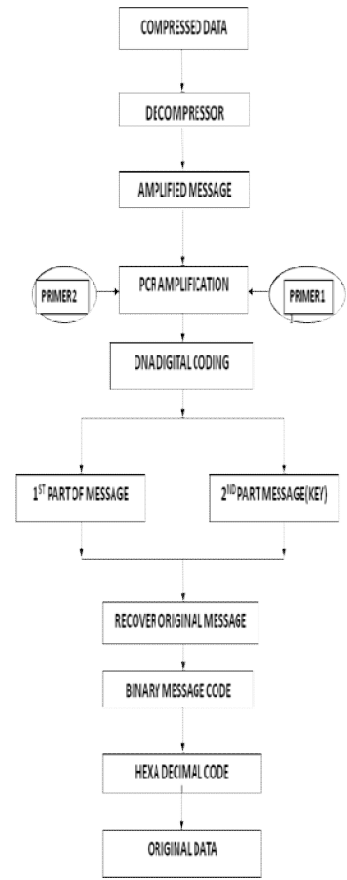

Fig3: DECRYPTION

## IX. SECURITY

Regardless of many differences between DNA and traditional cryptography, they both satisfy the same characteristic of cryptography. The security requirements should based depend only on the secrecy of decryption key. In this scheme encryption scheme as a key $k$A, $k$B and pair of PCR primer pairs.

The security of the encryption scheme come from two levels:

The first level is biological difficult problems its extremely difficult to recover the DNA digital code without knowing the correct two primer pairs. When the adversary gets the sending sample and tries pick out the message encoded sequence without knowing the correct PCR primer pairs would take 10^23 kinds of sequences. So only when both primer pairs are correct, then only retrieve the DNA digital code by amplification. On the other hand, if one the primer pair is wrong then it cause biological pollution, that leads to corruption of data.

PCR amplification require two primer pairs, to amplify the message encoded sequence , so we can use as one of the prime pairs as private key and another one as public key. It provides security unless there is development in DNA sciences , in case the DNA layer can be broken , then another layer is present to save the message from intruders as one part of message as key. So if an intruder break the DNA layer , the intruder can't get the full message because one part of message is kept as key. By combing the key with amplified binary message and key only, we recover original text else uncompleted message or polluted message. So,the biological and traditional method provides double safe guard to the message.

In case, intruder is hack our code, then we change the bit pattern in the DNA digital coding and also binary pattern, leads to Secure.

## X.CONCLUSION

In this paper, we designed Bi-serial DNA encryption algorithm containing technologies of DNA synthesis, PCR amplification, DNA digital coding, XOR operation as well as traditional cryptography.

The intended PCR two primer pairs was used as the key of this scheme that not independently designed by the sender or receiver. This operation could increase the security of encryption method. On the other hand, the traditional encryption method and DNA Digital Coding are used to preprocess operation we can get completely different cipher text from the same plaintext, which can effectively prevent attack from possible word as PCR primers. The complexity of biological difficult problem and cryptography computing difficulties provide a double layer security. And the security analysis shows that encryption scheme has high confidential strength. More over the cost of encryption will be greatly with process of techniques.

We have concentrated on the space and time complexity during computation, in that way introducing compression technique to compress the encrypted data and to minimize the utilization of memory and time during computation .This will provide potential support to our proposed system..

ACKNOWLEDGMENT

My sincere thanks to to my students (manicavel, jagadeesh & ragunath) who helped me to implement this project. And my friend Mr. M. Adimoolam a good hearted philanthropist. And a the lord almighty for his blessings.